\documentclass{webofc}
\usepackage[varg]{txfonts}
\usepackage[utf8]{inputenc}
\usepackage{lineno}
\usepackage{svg}
\usepackage{hyperref}
\usepackage{color}
\usepackage{hhline}
\usepackage{amsmath}
\usepackage{amssymb}
\usepackage{mathrsfs}
\usepackage{ bbold }
\usepackage{multirow} 
\usepackage{xspace}   
\usepackage{booktabs} 
\usepackage{paralist} 
\usepackage{tabularx}

\newcommand{\Hioo}{\texttt{H1oo}\xspace}

\newcommand{\orcid}[1]{\href{https://orcid.org/#1}{\includegraphics[width=10pt]{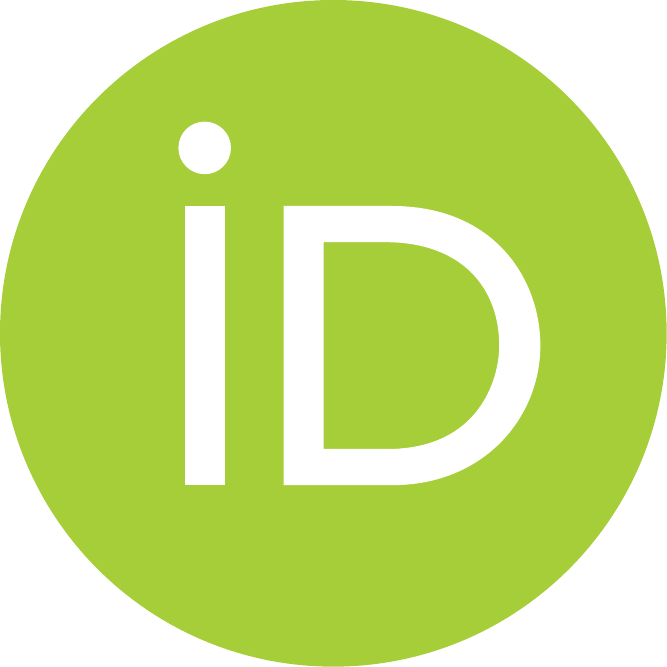}}}

\title{Preservation through modernisation: The software of the H1 experiment at HERA}
\date{\today}

\begin{document}

\author{
    \firstname{Daniel} \lastname{Britzger} \inst{1} \orcid{0000-0002-9246-7366} \and
    \firstname{Sergey} \lastname{Levonian} \inst{2} \orcid{0000-0002-4051-8694} \and
    \firstname{Stefan} \lastname{Schmitt} \inst{2} \orcid{0000-0001-8387-1853} \and
    \firstname{David} \lastname{South} \inst{2} \orcid{0000-0002-0786-6304}
    \\
    for the H1 Collaboration
}

\institute{
    Max-Planck-Institut für Physik, Föhringer Ring 6, 80805 M\"{u}nchen, Germany \and
    Deutsches Elektronen Synchrotron (DESY), Notkestr. 85, 22607 Hamburg, Germany
}

\abstract{
  The lepton--proton collisions produced at the HERA collider represent a unique high energy physics data set.
  A number of years after the end of collisions, the data collected by the H1 experiment, as well as the simulated events and all software needed for reconstruction, simulation and data analysis, were migrated into a preserved operational mode at DESY.
  A recent modernisation of the H1 software architecture has been performed, which will not only facilitate on going and future data analysis efforts with the new inclusion of modern analysis tools, but also ensure the long-term availability of the H1 data and associated software. 
  The present status of the H1 software stack, the data, simulations and the currently supported computing platforms for data analysis activities are discussed.
}

\maketitle

\section{The H1 experiment at HERA}
\label{sec:h1andhera}

Operating during the years $1992$ to $2007$, HERA at DESY is so far the only high energy lepton--proton ($ep$) collider in the world to have been constructed, where $27.6$~GeV electrons or positrons were brought into collision with $920$~GeV protons, resulting in a centre--of--mass energy of $319$~GeV.
The collision of point--like leptons with hadrons made HERA a unique tool for precise measurements of the structure of the proton.
Many other areas of particle physics were also accessible at HERA, including QCD and jets, heavy quark production, diffraction, electroweak physics, as well as the search for rare processes in $ep$ collisions.

The H1 detector~\cite{Abt:1996hi,Abt:1996xv} at HERA recorded the final state particles of $ep$ collision events, and features tracking detectors closest to the beam pipe, surrounded by electromagnetic and hadronic calorimetry, a muon system, and several further subdetector components.
Approximately $270,000$ readout channels were employed by the H1 detector.
A multi--level trigger system was employed to reduce the event-rate from the bunch crossing frequency of $96$~ns ($\approx 10$~MHz), and selected events were then stored with a rate of $20$--$50$\,Hz in the {\tt RAW} data format.
The total volume of {\tt RAW} $ep$ collision data recorded by the H1 detector and suitable for analysis amounts to about $75$~TB, and comprises approximately $1$ billion events collected in the years $1996$--$2007$.

Considering the planned Electron--Ion Collider in the US (EIC)~\cite{EIC}, the proposed Large Hadron-electron Collider at CERN (LHeC)~\cite{Agostini:2020fmq} and the proposed Electron-Ion Collider in China (EicC)~\cite{Anderle:2021wcy}, as well as many new related theoretical  developments, the unique $ep$ data from HERA retain their relevance for many years to come.
%

\section{The H1 software stack}
\label{sec:software}

The H1 Collaboration developed and maintains a sizeable software stack, which was used during data taking, and continues to be used not only for subsequent data reconstruction and high-level data analysis, but also for the simulation of the detector response to high energy physics (HEP) processes. 
The software stack is also crucial to perform physics analyses of the H1 $ep$ collision data.
The relevant components of the software stack are briefly described in the following.

\subsection{The core packages}
\label{sec:fortran}

The H1 core software is written almost entirely in {\tt FORTRAN 77}, was designed in a machine-independent manner\,\footnote{The H1 Collaboration employed many computing platforms and operating systems to analyse data, among them AIX, AILLANT, APOLLO, AXP, Ultrix, HPUX, IBM, IBMMVS, MAC, MIPS, OS-9, RTPC, VAC, VAXMVS, VMS, UNIX, SGI, SUN, LINUX, and most recently i686 and amd64 with RHEL or Scientific Linux distributions.}, and has served the experiment well since $1988$.
It is this software which creates the basic format, the Data Summary Tape ({\tt DST}) from the {\tt RAW} data, as well as providing the means to simulate $ep$ collision events for Monte Carlo (MC) comparisons to data.

The {\tt FORTRAN} software is based on two key components, {\tt BOS}~\cite{bos} and {\tt F-PACK}~\cite{fpack}, and is arranged in a series of strictly modular packages or modules, with self-contained sets of routines and clear input/output (I/O) interfaces.
{\tt BOS} is a dynamic memory management system for defined data blocks and their persistent I/O. The system supports a modular structure of the application program and portability for both the software and the data sets.
{\tt F-PACK} is a machine-independent general I/O package for data blocks. It performs automatic word format conversions for different machine representations (IBM, VAX, DEC, IEEE) and supports fast access to subsets of data through indexed files.
An important feature of {\tt F-PACK} is the support of keyed/ordered access files, which is essential for data-base-like applications.
The modules of the H1 packages communicate with each other strictly only via {\tt BOS} banks, and larger programs, such as the reconstruction program, consist of a plain series of module calls.

The H1 reconstruction package {\tt H1REC} comprises approximately $50$ different sub-modules for noise suppression, clustering, calibration, tracking, vertexing, tagging, for the combination of different sub-detector quantities, as well as electron identification and the reconstruction of deep inelastic scattering kinematic variables.
The detector simulation package {\tt H1SIM} is based on {\tt GEANT3}~\cite{geant3}.
It has a modular structure and contains the geometry definition of all sub-detectors, the magnetic field map, and tools for fast simulation and the shower library.
Taking the relevant run conditions from an {\tt Oracle} database, MC events are produced in the same format as the data, with some additional information.
The same reconstruction software as used on the data is then applied to the simulated MC events.

Further modules include the H1 global database {\tt H1NDB}, the data management system {\tt DATMAN}, the event filter modules {\tt H1L4} and {\tt H1ECLASS}, trigger simulation modules {\tt H1TRIG} and {\tt H1FTTEMU}, as well as libraries for useful numerical utilities {\tt H1UTIL} and utilities for MC production {\tt H1MCUTIL}.
Finally, the {\tt LOOK} package is a general system for graphics applications in physics, and uses {\tt GKS}~\cite{gks} for basic graphical operations with a system of utility subprograms to prepare own display programs. As such, {\tt LOOK} provides the core framework for the event display, {\tt H1ED}.

Only a few mandatory external dependencies exist, namely \texttt{CERNLIB}~\cite{cernlib}, \texttt{GEANT3}, \texttt{GKS}, \texttt{oracle-instant} and the {\tt FORTRAN} compiler, presently \texttt{gfortran}.
The first three packages are no longer maintained, but H1 retains a copy in its internal software stack.

\subsection{Analysis data and software: \Hioo}
\label{sec:h1oo}

In the year $2000$, a few years ahead of the HERA II phase of data taking, H1 made the decision to develop a new object--oriented analysis core framework in {\tt C++~}, \Hioo~\cite{Berthon:2000vw,Peez:2003iv,Katzy:865583,Steder:2011zz,Laycock:2012ps}.
The goal of the \Hioo project was to improve the overall efficiency in H1 physics analysis and performance by providing a modern analysis environment, new standardised and fast event selection facilities and a new, common data format.

The design and development principles employed in \Hioo were closely aligned to the ambitious developments of the emerging {\tt ROOT}~\cite{root} framework at CERN.
Concepts such as (multiple) inheritance, dynamic instantiation of objects via Names, consistent Setter and Getter functions for the use with the interactive {\tt C++} interpreter (CINT) and comprehensive source-code documentation are among the similarities in the design principles.
\Hioo also makes use of the generic collection of tools for high-performance I/O, as well as the interactive analysis environment and graphical opportunities provided by {\tt ROOT}.
The use of a global singleton class to provide convenient access to all relevant quantities is also very similar to the {\tt ROOT} paradigm.

The \Hioo framework is structured into about $50$ sub-packages and consists of more than $600$ {\tt C++}~classes, inheriting from the core {\tt ROOT} class {\tt TObject}.
The highly standardised data transfer methods between the sub-packages via {\tt BOS} and {\tt F-PACK} in the {\tt FORTRAN} software described in the previous section also provided a stable access interface to read the DST event files from \Hioo in {\tt C++}.
A more complete discussion of the structure of the \Hioo software can be found in~\cite{Steder:2011zz}.

For \Hioo, new data formats were designed to define new high-level analysis objects, such as jets, heavy-quark tagging information or (re-)calibrated particle candidates, with improved data I/O via {\tt TTree}s stored in {\tt ROOT} files.
The \Hioo data format typically used for analysis is in reality comprised of two persistent file formats, to be used in tandem: the "H1 Analysis Tag" ({\tt HAT}) which contains simple, calibrated variables to perform a fast event selection, and the larger "Micro Object Data Store" ({\tt mODS}) which contains additional information on identified particles.
A third \Hioo file format, the "Object Data Store" ({\tt ODS}), may be accessed transiently during an analysis event loop, and provides an interface to the full information stored on the original {\tt DST} file in {\tt ROOT} format.
This transient {\tt ODS} access is the only part of an \Hioo analysis which requires the {\tt FORTRAN} software.

The \Hioo framework provides many utilities for data analysis, wrappers for data access, templates for analysis codes, and standards for H1 physics analyses.
An event display {\tt H1Red} provides visualisation tools in the {\tt ROOT} analysis environment, and features full backward compatibility with {\tt LOOK} and {\tt H1ED}.

Whilst at some level each H1 physics analysis retains its own specific high-level analysis code and workflows, the common \Hioo framework greatly helped to standardise event selection, particle and event reconstruction, calibration, production of histograms and the assessment of measurement uncertainties and systematic errors.
This has had additional benefits in terms of shared analysis code, expert knowledge and working environments and, perhaps most importantly, data handling where members of the collaboration all use the same file formats.

\subsection{Software access, distribution and documentation}
\label{sec:softdoc}

All H1 software and source code are available to the members of the collaboration through their DESY credentials.
Pre-compiled executables, shared libraries and a working environment are provided centrally via the DESY-IT infrastructure for the supported operating system(s), which is currently {\tt CentOS7}.
The {\tt FORTRAN} code is documented in about $60$ internal H1 software notes that were written in a standardised, journal--like format.
Some packages were also presented and documented in journals or conference proceedings.
Each of the approximately $10\,000$ subroutines is stored in a separate file and includes comprehensive, standardised in--code documentation as well as a version history.


The \Hioo\ framework comes together with its own manual, tutorials, examples and a central, internal web--page.
Online {\tt HTML} code documentation is provided using {\tt ROOT}'s {\tt THtml} class and features more than $2000$ individual web--pages.
A single, central web--server provided by DESY-IT hosts all web-resources from all parts of the collaboration, such as meeting notes, hardware documentation, run--dependent documentation, analysis notes, trigger details, papers and so on.

\section{DPHEP and the data preservation model for H1}
\label{sec:earlydphep}

Data taking at HERA ended in June $2007$, and was soon followed by other high energy physics experiments of the same generation, namely BaBar at the PEP-II $e^{+}e^{-}$ collider at SLAC (April $2008$) and the D{\O} and CDF experiments at the Tevatron $p\bar{p}$ collider (September $2011$).
In order to perform a detailed evaluation of how to preserve high energy physics data for long--term analysis, an inter--experimental study group "Data Preservation in High Energy Physics" (DPHEP)~\cite{dphep-web} was formed at the end of 2008 to systematically investigate all technical and organisational aspects of this subject.
A series of six workshops took place between $2009$ and $2012$ and following a short interim report in $2009$~\cite{dpheppub1}, a full report was released in May $2012$~\cite{dpheppub2}.

The complete H1 {\tt RAW} collision data comprises around $75$~TB, the set of compressed \texttt{DST} data amounts to about $20$~TB and the analysis level \Hioo files are about $4$~TB.
Other data, such as random trigger streams, noise files, cosmic data, luminosity monitor and other calibration data amounts to a few TB.
The total volume of preserved simulated MC sets is around the same size of the data,
and hence the total volume of preserved data is about $0.5$~PB.
This is about the similar volume of {\tt RAW} data written out in just three days by the ATLAS detector at the LHC.

A key issue in data preservation is that in addition to the data themselves the associated software also needs to be considered, such as the programs for data access, reconstruction, simulation and analysis programs.
These programs also provide an important documentation of the data themselves.
In contrast, without a well defined and understood software environment the scientific potential of the data is limited, e.g.\ the possibility to study new observables or to incorporate new reconstruction algorithms, detector simulations or event generators may not be possible. 
With this is mind, a model for data preservation was devised~\cite{South:2011zp}, based on a series of levels of increasing complexity into which projects could be assigned~\cite{South:2012vf}, as shown in Table~\ref{tab:levels}.

\begin{table}[h]
\centering
\footnotesize
\caption{Data preservation levels defined by the DPHEP Study Group.}
\label{tab:levels}
\begin{tabular}{cll}
\toprule
Level &  Preservation Model & Use Case \\ \midrule
1 & Provide additional documentation &  Publication related info search \\
2 & Preserve the data in a simplified format & Outreach, simple training analyses and data exchange\\
3 & Preserve the analysis level software & Full scientific analysis possible, based on the \\
    & and the data format & existing reconstruction \\
4 & Preserve the reconstruction and simulation & Retain the full flexibility and potential of the \\
& software as well as the basic level data & experimental data \\ \bottomrule
\end{tabular}
\end{table}

As new experimental methods, new phenomenological models or new observables, are likely to be the prime reasons for analysing event data again, scenarios arise where only a more comprehensive preservation model will provide the necessary ingredients.
For example if a parameter in the reconstruction algorithm is kinematically limiting, or a new simulation written in a modern computing language provides only incompatible interfaces.
This approach assumes, justifiably, that it would be impossible to rewrite experiment specific software such as detector simulation or reconstruction code from scratch, due to missing expert knowledge about real hardware components and complexity of these programs.
With this in mind, H1 followed a DPHEP level~$4$ preservation model~\cite{South:2012vh}, and preserves the analysis level data formats, as well as the most basic level ({\tt RAW}) data, and all software.
It is clear that this level of preservation will necessarily include the full range of both experiment--specific and external software library dependencies.
However, the benefit of such a model is that the full physics analysis chain is available and full flexibility is retained for future use, similar to a real running experiment.

Whilst beyond the scope of this paper, it is also worth noting that a significant DPHEP level~$1$ preservation effort has also been performed.
This ensures all relevant digital and non--digital documentation concerning all aspects of H1 are safeguarded for future use~\cite{South:2012vh}.
A dedicated web server now hosts all of the documentation of the H1 experiment through static web pages, representing a single resource for knowledge transfer for the members of the collaboration.
Furthermore, a new, simpler operational model was also officially adopted by the collaboration in July $2012$ to ensure the efficient long term governance via the H1 Physics Board, which is made up of experts from all areas of the experiment.

\section{Preserved operational mode: 2012--2020}
\label{sec:2012-2020}

The final version of the reconstructed data, {\tt DST 7}, was produced in December $2010$ with the inclusion of the HERA I data from $1996$--$2000$.
This reprocessing campaign utilised a dedicated "dataflow meta--computing framework"~\cite{Campbell:2001chep}, which reached a stable performance of $60$ million events per day.
This represented a factor of $20$ improvement with respect to the previous architecture employed during earlier HERA I reprocessing campaigns~\cite{DagoretCampagne:1997wd}.
It was not until summer $2012$ that the HERA I \Hioo data and MC to be used for analysis were fully prepared including all relevant alignments and calibrations.
The \Hioo production release {\tt 4.0.25}, which uses {\tt ROOT5.34}, represented the end of major development, and was not only used to produce the data and MC for long term analysis but would also be the software release to be used for the remainder of the decade and beyond.

The H1 computing infrastructure included a dedicated $1200$ core batch system, significant storage capabilities comprising both tape and disk, and several large working group servers.
As part of the $2012$ transition into the new "preserved operational mode", the majority of these resources were phased out and replaced by a computing infrastructure centrally managed by DESY--IT and largely shared with other experiments.
The H1 data identified for preservation (see section~\ref{sec:earlydphep}) was relocated during $2014$--$2015$ to a dedicated, dCache~\cite{dcache} based storage at DESY, featuring two copies of the {\tt RAW} data on different tape media as well as an online pool for access to the most popular data.

As described in section~\ref{sec:earlydphep}, the H1 strategy is to retain the full flexibility of the data using a DPHEP level $4$ data preservation model.
This requires that the H1 software stack and all of its dependencies are continuously maintained and repeatedly updated and tested whenever a change in either the infrastructure or external dependencies is made.
These continuous migrations were made possible due to the modular structure, the high quality and stability of the packages, and H1 further benefited from the choice of stable programming languages for the software, {\tt FORTRAN 77}, {\tt C} and {\tt C++98}, and only a very moderate usage of {\tt shell} and {\tt perl} scripts.

The main OS employed by DESY--IT in $2012$ was $32$--bit Scientific Linux DESY 5 ({\tt SLD5}), and whilst this became the baseline version for H1, work was begun that year to migrate to $64$--bit, an important step to achieve given that next generation OS versions would only be available as $64$--bit versions.
{\tt SLD6} became the standard at DESY after $2015$.
Each OS migration revealed certain dependencies or required additional updates from DESY--IT, for example new versions of {\tt Oracle} or {\tt dCache}, which required only small changes to the H1 software to retain compatibility.
For example, after the transition from {\tt SLD6} to {\tt CentOS7} in $2020$, the library {\tt "libdcap.so"} used for direct IO access to {\tt dCache} systems
was removed from the DESY computing environment, resulting in corresponding, minor adjustments to the H1 software.

The use of external software in the \Hioo software was reduced as much as possible in $2012$ and whilst those remaining, namely {\tt ROOT}, {\tt fastjet}~\cite{Cacciari:2005hq} and {\tt neurobayes--expert}~\cite{Feindt:2006pm}, were migrated to the new operating systems, the last stable versions from 2012 were kept.
Table \ref{tab:2} shows a breakdown of the different components of the H1 software stack, and the status of the various dependencies during the preserved operational mode ($2012$--$2020$).

Whilst the migrations required only moderate person power, due to necessary validations and the overall complexity of the software stack as such they had to be performed by experienced H1 software experts.
To assist in this exercise a framework~\cite{Ozerov:2013isa} was developed, consisting of a series of well defined validation tests, to check for consistency every time a part of the software stack or environment was updated.
Whilst not fully deployed in an automated way, this project nevertheless showed its value 
during the first migrations to $64$--bit operating systems.

\begin{table}[ht]
  \centering
  \footnotesize
  \caption{Breakdown of the H1 software stack and its dependencies for the preserved operational mode ($2012$--$2020$).}
  \begin{tabular}{lcll}
    \toprule
    Component & Responsible & Maintained packages & Discontinued packages\\
    \midrule
    H1 software  &
    H1 &
    H1 core software, \Hioo & -- \\
    \addlinespace
    OS dependencies &
    DESY--IT &
    Oracle, dCache, web--services, & CVS \\
    (continuous updates) &
    &  compilers, GNU utilities, & \\
    &
    &
    gmake, system libraries & \\
    \addlinespace
    External dependencies &
    H1 &
    fastjet, neurobayes--expert, & CERNLIB, GKS, GEANT3, \\
    (selected fixed releases) &
    &
    MC generators & ROOT5, LHAPDF5, \\
    &
    &
    & MC generators \\
    \bottomrule
 \end{tabular}
\label{tab:2}
\end{table}

\section{The modernisation of H1 software and computing for the 2020s}
\label{sec:new}

\subsection{Revisiting the status of the H1 software}
\label{sec:newh1soft}

In $2020$ the status of the H1 software and the data analysis capabilities were again revisited.
Although the entire software was successfully migrated to {\tt CentOS7} and the full data analysis capability is retained, the overall status of the H1 software had a few shortcomings.
Some of these were introduced causally over the time period of the data preservation effort, and include the following:
\begin{compactitem}
  \item The programming languages and standards ({\tt C++98}) are unattractive for young people to learn and slow down new developments and data analysis efforts
  \item Outdated dependencies, such as {\tt ROOT5}, complicate the usage of modern data analysis techniques
  \item Modern tools have not in general been introduced
  \item No link to an externally maintained package repository is present, meaning that new packages had to be provided manually
  \item New dependencies may be incompatible, for example due to different compiler standards like {\tt C++20}, or different interfaces such as MC event generators providing an event record in newer data formats
  \item The compilation and maintenance of the packages still requires a profound and detailed knowledge about the specific structure of the H1 software, as well as some insight into the historic development
  \item No concise build instructions of the entire H1 software stack are available
  \item The H1 core packages are bound to the DESY--IT infrastructure and cannot be relocated
\end{compactitem}
\noindent
Some of these issues are very specific to the H1 software or computing infrastructure as it developed organically over three decades, but others are of a more general nature and may be repeated in data preservation efforts by other experiments in the future.
In particular for younger students, who want to perform a modern data analysis and also need to learn modern computing languages and techniques, the H1 software environment was not attractive.

\subsection{Restructuring the H1 software infrastructure after 10 years}
\label{sec:newh1inf}

At DESY, the only supported and locally deployed platform for the H1 software is {\tt CentOS7}, including all of the add--ons that it brings. 
However, in HEP the {\tt LCG} package repository~\cite{Pfeiffer:2004vkg,Roiser:2010zz} is commonly used as standard, and many recent dependencies, as well as modern compilers,
are provided by {\tt LCG/AA}~\cite{Hodgkins:2012cz} through {\tt cvmfs}~\cite{Buncic:2010zz} for {\tt CentOS7}.
H1 has decided to adopt this dependence, update all external dependencies accordingly, and to provide its software in a similar scheme to the members of the collaboration.

In a first instance, the platform {\tt x86\_64-centos7-gcc9-opt} and the {\tt LCG} version {\tt LCG\_97a} are selected for the next stable H1 software release. 
Many external packages are now provided through {\tt LCG}, such as {\tt ROOT}, {\tt fastjet}, {\tt neurobayes--expert}, thus reducing the maintenance cost to H1, as can be seen in Table~\ref{tab:3}.
This may be contrasted with the previous situation shown in Table~\ref{tab:2}.

\begin{table}[ht]
  \centering
  \footnotesize
 \caption{Breakdown of the H1 software stack and its dependencies from $2020$ onwards.} 
   \begin{tabular}{lcll}
    \toprule
    Component &  Responsible & Maintained packages & Discontinued packages \\
    \midrule
    H1 software &
    H1  &
    H1 core software, \Hioo & --\\
    \addlinespace
    OS dependencies &
    DESY--IT &
    Oracle, dCache, web--services, & -- \\
    (continuous updates) &
    &
    GNU utilities, git, & \\
    &
    &
    gmake, system libraries & \\
    \addlinespace
    External dependencies &
    H1 &
    -- & CERNLIB, GKS, GEANT3 \\
    (selected fixed releases) &
    &
    &
    (selected) MC generators \\
    \addlinespace
    External dependencies &
    LCG &
    LHADPF6, ROOT6, compilers, & -- \\
    (selected regular updates) &
    &
    fastjet, neurobayes--expert, & \\
    &
    &
    MC generators, (and as back up& \\
    &
    &
    option: Oracle, dCache, Git) & \\
    \bottomrule
 \end{tabular}
\label{tab:3}
\end{table}

The transition to the GNU compiler collection $9.2.0$, in contrast to the previously used version $4.8.5$, required a rebuild of all of the core {\tt FORTRAN} packages from scratch.
Due to the high quality of those packages, the stability of the {\tt FORTRAN 77} standard, and the many previous migrations of these codes, this transition was accomplished without complications, and only one case of memory corruption and few incorrect bit--wise logical operators needed to be fixed in about $950$k lines of code (LOC).
The discontinued packages \texttt{CERNLIB}, {\tt GKS}\,\footnote{GKS was obtained from \url{https://github.com/Starlink/starlink/tree/master/libraries/gks}} and {\tt GEANT3} could also be sourced.
%
%
The important low--level packages {\tt BOS} and {\tt FPACK}, which are needed for the interface between the H1 core packages and the access to the data, have been tested successfully.

More important for data analysis is the migration of the H1 analysis framework \Hioo, which is $300$k LOC written in {\tt C++98} and where the last stable version was built in $2012$ upon {\tt ROOT5.34}\,\footnote{\Hioo was initially developed using {\tt ROOT} version 2.}. 
The migration of \Hioo to {\tt ROOT6} was non--trivial, since for efficient data access, \Hioo makes extensive use of dynamic instantiation of named classes and dynamic scopes.
This functionality was provided in {\tt ROOT5} through {\tt CINT}, which was replaced by {\tt CLING} in {\tt ROOT6}.
Consequently, \Hioo\ had to strictly use the respective interfaces from {\tt ROOT6} and the direct access to {\tt CINT} functions had to be omitted.
A single class out of about 600 \Hioo\ classes could not be re--factored and was dropped, but this did not limit the usability of the analysis framework.
On the other hand, the new {\tt C++17} standard permits the implementation of convenient range--based for loops for \Hioo array classes, thus simplifying the code.
This feature facilitates the processing of particle lists in single events.
The object--oriented file formats {\tt mODS} and {\tt ODS} remain compatible, but a single object-oriented database file had to be regenerated.
Latest releases of {\tt fastjet}~v$3.3.2$ and {\tt neurobayes--expert}~v$3.7.0$ were integrated and successfully tested.

A complete release of all of the H1 software is provided to the collaboration in a single directory on {\tt afs}, which is mirrored to {\tt cvmfs} and the internal \texttt{nfs} shared file system at DESY.
This directory contains the binaries, libraries and the include files for all H1 software packages, together with compatible database snapshots.
A small setup script also initialises the compatible LCG release {\tt LCG\_97a/x86\_64-centos7-gcc9-opt}.
Compatible {\tt Oracle} and {\tt dCache} libraries are further provided from the {\tt LCG} mirror, and serve as a backup option if the DESY--IT libraries fail.

\subsection{Python--based and interactive data analysis with \Hioo}
\label{sec:python}

One great benefit from the close integration of \Hioo into the {\tt ROOT}--ecosystem is obtained from the {\tt PyROOT} developments in {\tt ROOT6}~\cite{Galli:2020boj}.
Through {\tt ROOT}'s automatised {\tt Python}--{\tt C++} bindings, the full functionality of \Hioo\ is available from {\tt Python}.
Together with the range--based operators in \Hioo, this enables a fully pythonic analysis of the H1 data and makes use of the \Hioo\ analysis framework.
Interactive data analyses in {\tt Python}, or using {\tt ROOT}'s {\tt C++} interpreter, are also possible, and may become a valuable option for future training or outreach activities.

Since \Hioo\ is provided through a global shared file system ({\tt cvmfs} or {\tt afs}), similar to the {\tt LCG} releases, it is fully relocatable and H1 data analysis or MC production can now be performed on any {\tt CentOS7} system anywhere in the world.
Alternatively, standardised {\tt CentOS7} containers, for example from {\tt CernVM}~\cite{Blomer:2017gpo}, can be employed on other platforms.
For a quick start for students or newcomers, a few new code examples are provided to H1 Collaboration members, both in {\tt C++} and {\tt Python}.

\subsection{Future maintenance of the H1 software stack}
\label{sec:future}

Future migrations of the H1 software will be connected to given versions of {\tt LCG} releases, featuring newer platforms, dependencies and compiler versions.
In order to facilitate this dependence a few additional developments were done.

All code repositories were migrated to {\tt Git}, which is the current standard for code management, an important transition given that most younger developers are unfamiliar with {\tt CVS} (or going even further back, {\tt CMZ}).
The DESY--IT central {\tt Git} repository hosting service {\tt Bitbucket}~\cite{stash}, which provides authentication for H1 members through their DESY credentials, replaces the old CVS web tools, thus reducing the maintenance cost for H1.

A new set of build instructions has been developed that allow to rebuild the entire H1 software stack from scratch.
Historic dependencies to central resources were removed and the build system was further simplified.
The build system remains dependent mainly on {\tt gmake}, which has proven to be very reliable over a long period of time.
Standardised software tests are currently under development.
About $50$ examples of analysis code are also kept in the {\tt Git} archive alongside the official H1 code, serving as a valuable reference for ongoing activities and for new data analysts.

\subsection{Singularity and containerised workflows}
\label{sec:containers}

The advent of containerisation and {\tt Singularity}~\cite{singularity} provides great opportunities for the preservation of data and software in HEP.
As binaries and libraries of the H1 software were retained in the current storage systems  deployed by DESY--IT on together with the complete set up procedures, these can now conveniently be used with {\tt Singularity} and compatible {\tt SLD5} or {\tt SLD6} containers.
Both {\tt SLD5} and {\tt SLD6} binaries have been tested and full functionality has been proven with standard containers from DESY--IT.
The use of {\tt Singularity} represents the retrospective realisation of the DPHEP level 3 `encapsulation' of the H1 software.

It is worth noting that whilst the use of containers and virtualisation offers potential solutions to data preservation, further development can only take place within the constraints of the original environment and technologies.
The modernisation program described in this paper has not only increased the longevity of the H1 software but undoubtedly made it more accessible and attractive to the next generation of collaborators.

\section{Summary and Outlook}
\label{sec:summary}

The H1 experiment recorded a unique data set of lepton--proton collisions in the years $1992$ to $2007$ and developed a sizeable software stack for its processing and analysis. 
The H1 Collaboration maintains these data, the related software and the documentation. 
Several data analysis activities are still ongoing and new analysis projects are beginning due to the uniqueness of this scientific data set and the increasing interest of the HEP community in $ep$ scattering.
This interest is also reflected in the recent addition of new collaboration members.
An additional $18$ papers ($8\%$ of the total) have been published by the H1 Collaboration since $2012$, which was already five years after the end of data taking.

All core software packages that were developed in the years $1988$ to $2012$ have been successfully migrated to a modern computing platform (\texttt{amd64} (x86-64), {\tt CentOS7}, {\tt GNU}--compilers $9.2$).
These software modules are required for data access, data processing, reconstruction, simulation, visualisation and, of course, high--level data analysis.
External dependencies were updated to latest releases and are now obtained from the \texttt{LCG/AA} package repository.
All other IT services are hosted by DESY--IT.

The common object--oriented data analysis framework \Hioo is now based on {\tt ROOT6} and supports the {\tt C++17} standard. 
This framework facilitates the communication within the collaboration members, provides a common standard, and additionally provides highly valuable documentation.
Some example programs and a few selected full high--level analysis codes from previous publications are prepared for newcomers.
The \Hioo analysis framework is now fully accessible in {\tt Python} and an online code documentation is available.
\Hioo now enables interactive analysis in {\tt C++} through {\tt CLING} or more commonly in {\tt Python}.
Container solutions have been implemented for backward compatibility and software tests.

The programs and libraries are provided to the members of the H1 Collaboration on shared global file systems for convenience.
All H1 software packages are now maintained in {\tt Git} and new build instructions for a complete re--build of the entire software stack have been prepared.
With these modifications to the H1 software architecture, H1 is confident to provide high quality data analysis capability of the unique HERA data in the future, using modern analysis tools, recent programming languages and on state--of--the--art platforms.

\section*{Acknowledgments}
The authors would like to thank all collaboration members, past and present, as well as our colleagues in the DESY--IT department who helped make and continue to make the H1 experiment and DESY a uniquely enjoyable experience. Credit is also due to those involved in the early days of the DPHEP Study Group. We would particularly like to recognise and acknowledge the tremendous amount of work done by our late friend and colleague Vitaliy Dodonov, whose care, dedication and attention to detail made it possible to migrate the H1 software to $64$--bit {\tt SLD6}, laying the groundwork for the later migration to {\tt CentOS7} in $2020$.



\begin{thebibliography}{10}
\footnotesize
\bibitem{Abt:1996hi}
I.~Abt \emph{et~al.},
{The H1 detector at HERA},
Nucl.\ Instrum.\ Meth.\ A \textbf{386} (1997) 310

\bibitem{Abt:1996xv}
I.~Abt \emph{et~al.},
{The tracking, calorimeter and muon detectors of the H1 experiment at HERA},
Nucl.\ Instrum.\ Meth.\ A \textbf{386} (1997) 348

\bibitem{EIC}
A.~Accardi  \emph{et~al.},
{Electron Ion Collider: The next QCD frontier},
Eur.\ Phys.\ J.\ A \textbf{52} (2016) 268
  
\bibitem{Agostini:2020fmq}
P.~Agostini \emph{et~al.},
{The Large Hadron--Electron Collider at the HL--LHC},
\href{https://arxiv.org/abs/2007.14491}{arXiv:2007.14491}
  
\bibitem{Anderle:2021wcy}
D.~Anderle \emph{et~al.},
{Electron--Ion Collider in China},
\href{https://arxiv.org/abs/2102.09222}{arXiv:2102.09222}

\bibitem{bos}
V.~Blobel,
{BOS and related packages},
Proc. 14th Workshop of the INFN Eloisatron Project; Data Structures for Particle Physics Experiments: Evolution or Revolution, Erice, Italy (1990), p.\ 1--6

\bibitem{fpack}
V.~Blobel,
{The F--package for input/output},
Proc. 6th Int. Conf. on Computing in High Energy Physics (CHEP 1992),
Annecy, France (1992), p.\ 755--758

\bibitem{geant3}
R.~Brun \emph{et al.},
{GEANT 3},
CERN DD/EE 84-1 (1984)

\bibitem{gks}
D.~A.~Duce,
{The graphical kernel system (GKS) ISO 7942},
Computer Standards \& Interfaces, Vol. 6, Issue {\bf 2} (1987) p.\ 235--237,
DOI: \href{https://doi.org/10.1016/0920-5489(87)90065-1}{10.1016/0920-5489(87)90065-1}

\bibitem{cernlib}
\emph{CERN Program Library, CERNLIB}, URL: \url{https://cernlib.web.cern.ch/cernlib/} [accessed 2021-02-24]

\bibitem{Berthon:2000vw}
U.~Berthon \emph{et al.},
{New data analysis environment in H1},
Proc. 11th Int. Conf. on Computing in High Energy and Nuclear Physics (CHEP 2000), Padua, Italy (2000), p.\ 700--703

\bibitem{Peez:2003iv}
M.~Peez,
{The new object oriented analysis framework for H1},
Proc. 13th Int. Conf. on Computing in High Energy and Nuclear Physics (CHEP 2003), La Jolla, USA (2003),
\href{https://arxiv.org/abs/physics/0306124}{arXiv:physics/0306124}

\bibitem{Katzy:865583}
J.~Katzy,
{H1OO: An analysis framework for H1},
Proc. 14th Int. Conf. on Computing in High Energy and Nuclear Physics (CHEP 2005), Interlaken, Switzerland (2004),
DOI: \href{https://doi.org/10.5170/CERN-2005-002.265}{10.5170/CERN-2005-002.265}

\bibitem{Steder:2011zz}
M.~Steder,
{H1OO: A centralised analysis framework for the H1 experiment},
Proc. 18th Int. Conf. on Computing in High Energy and Nuclear Physics (CHEP 2010),
Taipei, Taiwan (2010),
J.\ Phys.:\ Conf.\ Ser. \textbf{331} 032051 (2011),
DOI: \href{https://doi.org/10.1088/1742-6596/331/3/032051}{10.1088/1742-6596/331/3/032051}
 
\bibitem{Laycock:2012ps}
P.~Laycock,
{Ten years of object--oriented analysis on H1},
Proc. 14th Int. Workshop on Advanced Computing and Analysis Techniques in Physics Research (ACAT 2011), Uxbridge, UK (2011),
J.\ Phys.:\ Conf.\ Ser. \textbf{368} 012048 (2012),
DOI: \href{https://doi.org/10.1088/1742-6596/368/1/012048}{10.1088/1742-6596/368/1/012048}

\bibitem{root}
\emph{ROOT: Data Analysis Framework}, URL: \url{https://root.cern/} [accessed 2021-02-24]

\bibitem{dphep-web}
\emph{Data Preservation in High Energy Physics, DPHEP}, URL: \url{https://dphep.org/} [accessed 2021-02-24]

\bibitem{dpheppub1}
D.~Asner {\it et al.} [DPHEP Study Group],
{Data preservation in high energy physics},
\href{https://arxiv.org/abs/0912.0255}{arXiv:0912.0255}

\bibitem{dpheppub2}
Z.~Akopov {\it et al.} [DPHEP Study Group],
{Status report of the DPHEP Study Group: Towards a global effort for sustainable data preservation in high energy physics},
\href{https://arxiv.org/abs/1205.4667}{arXiv:1205.4667}

\bibitem{South:2011zp}
D.~M.~South,
{Data preservation in high energy physics},
Proc. 18th Int. Conf. on Computing in High Energy and Nuclear Physics (CHEP 2010),
Taipei, Taiwan (2010),
J.\ Phys.:\ Conf.\ Ser. \textbf{331} 012005 (2011),
DOI: \href{https://doi.org/10.1088/1742-6596/331/1/012005}{10.1088/1742-6596/331/1/012005}

\bibitem{South:2012vf}
D.~M.~South,
{Data preservation and long term analysis in high energy physics},
Proc. 19th Int. Conf. on Computing in High Energy and Nuclear Physics (CHEP 2012) New York, USA (2012),
J.\ Phys.:\ Conf.\ Ser. \textbf{396} 062018 (2012),
DOI: \href{https://doi.org/10.1088/1742-6596/396/6/062018}{10.1088/1742-6596/396/6/062018}

\bibitem{South:2012vh}
D.~M.~South and M.~Steder,
{The H1 data preservation project},
Proc. 19th Int. Conf. on Computing in High Energy and Nuclear Physics (CHEP 2012) New York, USA (2012),
J.\ Phys.:\ Conf.\ Ser. \textbf{396} 062019 (2012),
DOI: \href{https://doi.org/10.1088/1742-6596/396/6/062019}{10.1088/1742-6596/396/6/062019}

\bibitem{Campbell:2001chep}
A.~Campbell \emph{et al.},
{A dataflow meta--computing framework for event processing in the H1 experiment}, 
Proc. 12th Int. Conf. on Computing in High Energy and Nuclear Physics (CHEP 2001), Bejing, China (2001), p.\ 651--655

\bibitem{DagoretCampagne:1997wd}
S.~Dagoret-Campagne \emph{et al.},
{Reprocessing H1 data on the IN2P3 computer farm in $1997$},
Proc. 9th Int. Conf. on Computing in High Energy and Nuclear Physics (CHEP 1997), Berlin, Germany (1997).

\bibitem{dcache}
\emph{dCache: Distributed Storage for Scientific Data}, URL: \url{https://dcache.org} [accessed 2021-02-24]

\bibitem{Cacciari:2005hq}
M.~Cacciari and G.~P.~Salam,
{Dispelling the $N^{3}$ myth for the $k_t$ jet--finder},
Phys.\ Lett.\ B \textbf{641} 57 (2006),
DOI: \href{https://doi.org/10.1016/j.physletb.2006.08.037}{10.1016/j.physletb.2006.08.037}

\bibitem{Feindt:2006pm}
M.~Feindt and U.~Kerzel,
{The NeuroBayes neural network package},
Proc. 10th Int. Workshop on Advanced Computing and Analysis Techniques in Physics Research (ACAT 2005), Zeuthen, Germany (2005),
Nucl.\ Instrum.\ Meth.\ A \textbf{559} 190 (2006),
DOI: \href{https://doi.org/10.1016/j.nima.2005.11.166}{10.1016/j.nima.2005.11.166}

\bibitem{Ozerov:2013isa}
D.~Ozerov and D.~M.~South,
{A validation framework for the long term preservation of high energy physics data},
Proc. 20th Int. Conf. on Computing in High Energy and Nuclear Physics (CHEP 2013) Amsterdam, Netherlands (2013),
J.\ Phys.:\ Conf.\ Ser. \textbf{513} 042043 (2014),
DOI: \href{https://doi.org/10.1088/1742-6596/513/4/042043}{10.1088/1742-6596/513/4/042043}

\bibitem{Pfeiffer:2004vkg}
A.~Pfeiffer,
{Overview of the LCG applications area software projects},
Proc. IEEE Nuclear Science Symposium and Medical Imaging Conference (NSS/MIC 2004), Rome, Italy (2004),
DOI: \href{https://doi.org/10.1109/NSSMIC.2004.1462660}{10.1109/NSSMIC.2004.1462660}
 
\bibitem{Roiser:2010zz}
S.~Roiser, A.~Gaspar, Y.~Perrin and K.~Kruzelecki,
{Servicing HEP experiments with a complete set of ready integrated and configured common software components},
Proc. 17th Int. Conf. on Computing in High Energy and Nuclear Physics (CHEP 2009), Prague, Czech Republic (2009),
J.\ Phys.:\ Conf.\ Ser. \textbf{219} 042022 (2010),
DOI: \href{https://doi.org/10.1088/1742-6596/219/4/042022}{10.1088/1742-6596/219/4/042022}

\bibitem{Hodgkins:2012cz}
A.~L.~Hodgkins, V.~Diez and B.~Hegner,
{LCG/AA build infrastructure},
J.\ Phys.:\ Conf.\ Ser. \textbf{396}, 052026 (2012),
DOI: \href{https://doi.org/10.1088/1742-6596/396/5/052026}{10.1088/1742-6596/396/5/052026}

\bibitem{Buncic:2010zz}
P.~Buncic \emph{et al.},
{CernVM: A virtual software appliance for LHC applications},
Proc. 17th Int. Conf. on Computing in High Energy and Nuclear Physics (CHEP 2009), Prague, Czech Republic (2009),
J.\ Phys.:\ Conf.\ Ser. \textbf{219} 042003 (2010),
DOI: \href{https://doi.org/10.1088/1742-6596/219/4/042003}{10.1088/1742-6596/219/4/042003}

\bibitem{Galli:2020boj}
M.~Galli, E.~Tejedor, and S.~Wunsch,
{A new PyROOT: Modern, interoperable and more pythonic},
Proc. 24th Int. Conf. on Computing in High Energy and Nuclear Physics (CHEP 2019), Adelaide, Australia (2019),
EPJ\ Web\ Conf. \textbf{245} 06004 (2020),
DOI: \href{https://doi.org/10.1051/epjconf/202024506004}{10.1051/epjconf/202024506004}

\bibitem{Blomer:2017gpo}
J.~Blomer \emph{et al.}, 
{New directions in the CernVM file system},
{Proc. 22nd Int. Conf. on Computing in High Energy and Nuclear Physics (CHEP 2016)}, San Francisco, USA, (2016),
J.\ Phys.:\ Conf.\ Ser. {\bf 898} (2017) 062031,
DOI: \href{https://doi.org/10.1088/1742-6596/898/6/062031}{10.1088/1742-6596/898/6/062031}
  
\bibitem{stash}
\emph{DESY Bitbucket repository}, URL: \url{https://stash.desy.de/} [accessed 2021-02-24]

\bibitem{singularity}
G.~M.~Kurtzer, V.~Sochat and M.~W.~Bauer,
{Singularity: Scientific containers for mobility of compute},
PLoS ONE \textbf{12(5)} e0177459 (2017),
DOI: \href{https://doi.org/10.1371/journal.pone.0177459}{10.1371/journal.pone.0177459}

\end{thebibliography}
\end{document}